\documentclass[11pt]{article}
\newcommand{\Var}{\mbox{Var}}

\usepackage[mathscr]{eucal}
\usepackage{amsmath}
\usepackage[pctex32]{graphicx}
\usepackage{verbatim}
\usepackage{color}
\usepackage[usenames,dvipsnames]{xcolor}
\usepackage[round]{natbib}
\usepackage{amssymb,amsmath,amscd,amsthm,amsfonts}
\usepackage{latexsym}
\usepackage{lscape}
\usepackage{rotating}
\newtheorem{theorem}{Theorem}

\usepackage[T1]{fontenc}
\usepackage[utf8]{inputenc}
\usepackage{authblk}
\usepackage{chngpage}

\title{Extending the Archimedean copula methodology to model multivariate survival data grouped in clusters of variable size}
\author[1]{Leen Prenen}
\author[1]{Roel Braekers \thanks{Address: Roel Braekers, Interuniversity Institute for Biostatistics and statistical Bioinformatics, Universiteit Hasselt, Martelarenlaan 42, B-3500 Hasselt, Belgium. E-mail: roel.braekers@uhasselt.be}}
\author[2]{Luc Duchateau}
\affil[1]{Interuniversity Institute for Biostatistics and statistical Bioinformatics, Universiteit Hasselt, Belgium}
\affil[2]{Department of Physiology and Biometrics, Universiteit Gent, Belgium}
\date{}

\begin{document}
  \maketitle

\footnotesize

\section*{Abstract}

For the analysis of clustered survival data, two different types of models that take the association into account, are commonly used: frailty models and copula models. Frailty models assume that conditional on a frailty term for each cluster, the hazard functions of individuals within that cluster are independent. These unknown frailty terms with their imposed distribution are used to express the association between the different individuals in a cluster. Copula models on the other hand assume that the joint survival function of the individuals within a cluster is given by a copula function, evaluated in the marginal survival function of each individual. It is the copula function which describes the association between the lifetimes within a cluster. A major disadvantage of the present copula models over the frailty models is that the size of the different clusters must be small and equal in order to set up manageable estimation procedures for the different model parameters. We describe in this manuscript a copula model for clustered survival data where the clusters are allowed to be moderate to large and varying in size by considering the class of Archimedean copulas with completely monotone generator. We develop both one- and two-stage estimators for the different copula parameters. Furthermore we show the consistency and asymptotic normality of these estimators. Finally, we perform a simulation study to investigate the finite sample properties of the estimators. We illustrate the method on a data set containing the time to first insemination in cows, with cows clustered in herds.\\\\
\textbf{Keywords:} Archimedean copula, multivariate survival data, varying cluster size

\newpage

 \normalsize
\section{Introduction}

Multivariate survival data consist of multiple lifetimes which are linked to each other in some sense. In clustered survival data, subjects in the same cluster are assumed to share some characteristic or environment, and are therefore expected to be more similar with respect to the hazard of the event. For example, in a multi-center clinical trial, patients of one center form a separate cluster. To analyze this type of multivariate survival data, two different techniques that take the association between the individuals into account, are commonly used. On the one hand, frailty models are considered in which the hazard function of an individual within a cluster is investigated, conditional on an unknown common frailty term for this cluster. This approach is explained in detail in \citet{DuchateauJanssen2008} and \citet{Wienke2011}. On the other hand, in copula models, the joint survival function for all individuals within a cluster is modelled by a copula function which is evaluated in the marginal survival function of each individual. In copula models, the behaviour of each separate lifetime is investigated in combination with a copula function that controls for the association structure between the different lifetimes. \citet{ShihLouis1995} introduced the copula model first and provided estimation methods for the unknown parameters in a bivariate setting. \citet{Glidden2000} and  \citet{Andersen2005} extended the approach of \citet{ShihLouis1995} by introducing covariates in the marginal survival function. \citet{MassonnetJanssenDuchateau2009} extended these models further for clusters of size 4.

A major drawback of the reported techniques is that copula models are only used for clustered designs in which the cluster size is small and constant. For example, \citet{ShihLouis1995}, \citet{Glidden2000} and \citet{Andersen2005} considered clusters of size two while \citet{MassonnetJanssenDuchateau2009} modelled the time until infection in the four different quarters of a cow udder. Although \citet{Glidden2000} gives theoretical results for the Clayton copula in a balanced design with a fixed cluster size $N$ and \citet{OthusLi2010} do the same in an unbalanced design for the Gaussian copula model, to our knowledge, Archimedean copula models in general have not been used for clustered multivariate survival data with a cluster size of more than 4 or for a cluster size which differs over the clusters. The choice of a small and constant cluster size is a direct consequence of the difficulty to write down the likelihood function for the observed clustered survival data. For example, if the cluster size is equal to two, there are 4 different contributions to the likelihood for the observed outcomes within the cluster, depending on whether none, the first, the second or both individuals in this cluster are censored. This leads to a likelihood function consisting of 4 different terms where every term is found by taking derivatives of the joint survival function over the uncensored components in an observed couple. If the cluster size is three, the number of possible combinations increases to 8, while a cluster size of 4 leads to 16 different combinations. In a general setting with a cluster size equal to $n$, we have $2^n$ possible combinations. Since a likelihood function then also contains $2^n$ different possible terms and each term is found by taking derivatives of the joint survival function over the uncensored components in a combination, it is a huge task if a general $n$-dimensional copula function is considered for the association between the different individuals within a cluster. It is in practice impossible to calculate a closed form for all the derivatives of a copula function if the order $n$ is large. In frailty models such problems do not exist since it is assumed that conditional on a common frailty term, the individuals within a cluster are independent. The construction of the likelihood function for the frailty model uses this assumption by first looking at the conditional contribution of an individual within a cluster to the likelihood function by incorporating a frailty term and afterwards integrating over the frailty distribution. The frailty model approach allows that the number of individuals within a cluster varies. For the class of Archimedean copula functions the joint survival function can be rewritten as a mixture distribution of independent contributions as is the case in the frailty model approach. We show that this simplifies the construction of the likelihood function considerably and allows the cluster size to be moderate to large and varying.

The article is organized as follows. In Section \ref{description} we introduce a new formulation of the copula model by rewriting the likelihood contributions in terms of Laplace transforms. In Section \ref{estimation} we present the theoretical results concerning estimators arising from this model, starting from parametric and semiparametric approaches. Section \ref{distributions} gives an overview of a large class of distributions for which the likelihood contributions are easy to generate. In Sections \ref{dataexample} and \ref{simulation}, we report results for a data example along with some simulation results. Proofs of asymptotic results are given in the Appendix.

\section{Description of the model}\label{description}

We develop a copula model for clustered survival data in which the size of each cluster may be different. Let $K$ be the number of clusters ($i=1,\ldots,K$). In each cluster, we denote the lifetime for the different individuals by a positive random variable $T_{ij}$, $j=1,\ldots,n_i$ where $n_i$ is the number of individuals in cluster $i$. For each individual, we assume that there is an independent random censoring variable $C_{ij}$ such that under a right censoring scheme, the observed quantities are given by
$$ \begin{array}{l} X_{ij} = \min(T_{ij},C_{ij})\\
\delta_{ij}=I(T_{ij} \le C_{ij}) \end{array} \quad,i=1,\ldots,K, \quad j=1,\ldots,n_i.$$
The risk of failure may also depend on a set of covariates $\boldsymbol{Z}_{ij}$, which are possibly time-varying. We assume that the joint survival function for the lifetime of the different individuals within cluster $i$ is given by
\begin{eqnarray*}
S(t_{i1},\dots,t_{i n_i}|\boldsymbol{Z}_{i1},\dots,\boldsymbol{Z}_{in_i}) &=& P(T_{i1}>t_{i1},\ldots,T_{i n_i}> t_{i n_i}|\boldsymbol{Z}_{i1},\dots,\boldsymbol{Z}_{in_i})\\
&=&\varphi_\theta\left[\varphi_\theta^{-1}\left(S(t_{i1}|\boldsymbol{Z}_{i1})\right) + \dots + \varphi_\theta^{-1}\left(S(t_{i n_i}|\boldsymbol{Z}_{i n_i})\right)\right]
\end{eqnarray*}
where $S(t_{ij}|\boldsymbol{Z}_{ij})=P(T_{ij}>t_{ij}|\boldsymbol{Z}_{ij})$ is a common marginal survival model for the lifetime $T_{ij}$, given $\boldsymbol{Z}_{ij}$. The generator $\varphi_\theta: [0,\infty[ \to [0,1]$ of a parametric Archimedean copula family is a continuous strictly decreasing function with $\varphi_\theta(0)=1$ and $\varphi_\theta(\infty)=0$. We denote by $\varphi_\theta^{-1}$ the inverse function of $\varphi_\theta$. Since we want the Archimedean copula function to be correctly defined for any cluster size, we assume that this generator is completely monotonic. This means that all the derivatives exist and have alternating signs: $(-1)^m\frac{d^m}{dt^m} \varphi_\theta (t) \ge 0$, for all $t>0$ and $m=0,1,2,\ldots$ (see \citet{Nelsen2006}). The generator $\varphi_\theta$ is a Laplace transformation of a positive distribution function $G_\theta(x)$ with $\bar{G}_\theta(0)=1$ \citep{Joe1997},
$$\varphi_\theta(t)=\int\limits_0^{+\infty}e^{-tx}dG_{\theta}(x), \quad t\ge 0.$$
Hence we can rewrite the joint survival function for cluster $i$ as
\begin{eqnarray}
S(t_{i1},\dots,t_{i n_i}|\boldsymbol{Z}_{i1},\dots,\boldsymbol{Z}_{i n_i}) & = \int\limits_0^{+\infty} e^{-x \sum\limits_{j=1}^{n_i} \varphi_\theta^{-1}\left(S(t_{ij}|\boldsymbol{Z}_{ij})\right)} dG_\theta(x)\label{eq1}\\
& = \int\limits_0^{+\infty} \prod\limits_{j=1}^{n_i}e^{-x \varphi_\theta^{-1}\left(S(t_{ij}|\boldsymbol{Z}_{ij})\right)} dG_\theta(x).\nonumber
\end{eqnarray}
Note that we find a similar expression for the joint survival function in frailty models \citep[p.119]{DuchateauJanssen2008}, with $G_\theta(x)$ the frailty distribution of the unknown frailty term in the cluster. Mimicking this idea, the Archimedean copula function can be seen as a mixture distribution, consisting of independent and identically distributed components which each depend on a common factor that has $G_\theta$ as distribution. We use this analogy to derive the likelihood function. The contribution of cluster $i$, with cluster size $n_i$, to the likelihood function corresponds to the derivative of the $n_i$-dimensional joint survival function over all uncensored individuals in this cluster. The joint survival function does not change when the individuals within the cluster are permuted. Consequently, only the number of uncensored individuals determines the derivative. Hence, the contribution of cluster $i$ to the likelihood function is given by
$$L_i = (-1)^{d_{i}} \frac{\partial^{d_i}}{\partial\{\delta_{ij}=1\}} S(x_{i1},\ldots,x_{i n_i}|\boldsymbol{Z}_{i1},\dots,\boldsymbol{Z}_{i n_i})$$
where $\partial\{\delta_{ij}=1\}$ is the set of uncensored individuals in cluster $i$ and $d_i=\sum\limits_{j=1}^{n_i} \delta_{ij}$, the size of this set.

Using representation (\ref{eq1}) of the joint survival function, this derivative is given by
\begin{eqnarray*}
L_i
&=&\int\limits_0^{+\infty} e^{-x \sum\limits_{j=1}^{n_i} \varphi^{-1}_\theta(S(x_{ij}|\boldsymbol{Z}_{ij}))} \prod\limits_{j=1}^{n_i} \left[ \frac{-x f(x_{ij}|\boldsymbol{Z}_{ij})}{\varphi'_\theta (\varphi_\theta^{-1}(S(x_{ij}|\boldsymbol{Z}_{ij})))} \right]^{\delta_{ij}} dG_\theta(x)
\end{eqnarray*}
where $f=-dS/dt$ is the conditional density of the lifetime $X_{ij}$.

Combining the contributions over the different clusters, we get the following likelihood function
\begin{eqnarray}
L&=& \prod_{i=1}^K \int\limits_0^{+\infty} e^{-x \sum\limits_{j=1}^{n_i} \varphi_{\theta}^{-1}(S(x_{ij}|\boldsymbol{Z}_{ij}))} \prod\limits_{j=1}^{n_i}
\left[\frac{-x f(x_{ij}|\boldsymbol{Z_{ij}})}{\varphi_\theta'(\varphi_\theta^{-1}(S(x_{ij}|\boldsymbol{Z_{ij}})))}\right]^{\delta_{ij}}
dG_{\theta}(x) \nonumber\\
&=& \prod_{i=1}^K \int\limits_0^{+\infty} \prod_{j=1}^{n_i}e^{-x\varphi_\theta^{-1}(S(x_{ij}|\boldsymbol{Z}_{ij}))}
\left[\frac{-x f(x_{ij}|\boldsymbol{Z}_{ij})}{\varphi_\theta'(\varphi_\theta^{-1}(S(x_{ij}|\boldsymbol{Z_{ij}})))}\right]^{\delta_{ij}}
dG_{\theta}(x).\label{eq2}
\end{eqnarray}
In general it is difficult to evaluate expression (\ref{eq2}) except for very specific choices of the distribution $G_{\theta}$. Since the generator $\varphi_\theta$ is the Laplace transform of $G_\theta$, there is an alternative expression for this likelihood function which is found by using derivatives of this generator, i.e. $\varphi_\theta^{(m)}(t)=\int\limits_0^{+\infty} (-x)^m e^{-tx} dG_\theta (x)$. Hence the likelihood function can be rewritten as
\begin{equation}
L=\prod\limits_{i=1}^K \left( \prod\limits_{j=1}^{n_i} \left[\frac{f(x_{ij}|\boldsymbol{Z}_{ij})}{\varphi_\theta'(\varphi_\theta^{-1}(S(x_{ij}|\boldsymbol{Z}_{ij})))}\right]^{\delta_{ij}}\right)
\varphi_\theta^{(d_i)}\left(\sum\limits_{j=1}^{n_i}\varphi_\theta^{-1}(S(x_{ij}|\boldsymbol{Z}_{ij}))\right).\label{eq3}
\end{equation}

\section{The estimation procedures}\label{estimation}

We investigate the one-stage parametric estimation and two-stage parametric and semi-parametric estimation. \citet{DuchateauJanssen2008} demonstrate how this can be done for a bivariate survival data set, while \citet{ShihLouis1995} derive asymptotic properties of the estimators. \citet{Joe1997, Joe2005} discusses a general framework for studying asymptotic efficiency. We extend their results to clustered survival data with clusters of varying and possibly large size.

For equal-sized clusters with cluster size $n$ having the same covariate structure, baseline survival functions can be estimated for each $j^\text{th}$ univariate margin, $j=1,\dots,n$, where the $j^\text{th}$ subject always has the same covariate information. Since in our application clusters have varying size, we cannot order the components in a cluster and then estimate the baseline survival of all $j$ components. We assume that all subjects have the same baseline survival, whatever the cluster, and introduce subject specific covariate information.

\subsection{One-stage parametric estimation}
Let $\boldsymbol{\beta}$ be the parameter vector for the margins, containing distribution-specific parameters for the baseline survival and covariate effects. We use the likelihood function $L(\boldsymbol{\beta},\theta)$ as derived in (\ref{eq2}) and (\ref{eq3}).
Write $\mathbf{U}_{\boldsymbol{\beta}}(\boldsymbol{\beta},\theta)=\frac{\partial \log L(\boldsymbol{\beta},\theta)}{\partial \boldsymbol{\beta}},U_{\theta}(\boldsymbol{\beta},\theta) =  \frac{\partial \log L(\boldsymbol{\beta},\theta)}{\partial \theta}$. Solving
$$\left\{ \begin{array}{l}
\mathbf{U}_{\boldsymbol{\beta}}(\boldsymbol{\beta},\theta)=0\\
U_{\theta}(\boldsymbol{\beta},\theta)=0
\end{array} \right.$$
simultaneously, we find the maximum likelihood estimate $(\hat{\boldsymbol{\beta}},\hat{\theta})$. From maximum likelihood theory \citep{CoxHinkley1974}, we know that under regularity conditions, $\sqrt{K}(\hat{\boldsymbol{\beta}}-\boldsymbol{\beta},\hat{\theta}-\theta)$ converges to a multivariate normal distribution with mean vector zero and variance-covariance matrix $\mathbf{I}^{-1}$, where $\mathbf{I}$ is partitioned into blocks:
$$\mathbf{I}=\left(\begin{array}{c c} \mathbf{I}_{\boldsymbol{\beta}\boldsymbol{\beta}} & \mathbf{I}_{\boldsymbol{\beta}\theta}\\
\mathbf{I}_{\theta\boldsymbol{\beta}} & I_{\theta\theta}\end{array}\right).$$
Here, $K \mathbf{I}_{\boldsymbol{\beta}\boldsymbol{\beta}}$ is the variance-covariance matrix of $\mathbf{U}_{\boldsymbol{\beta}}$, $K \mathbf{I}_{\boldsymbol{\beta}\theta}$ is the covariance vector between $\mathbf{U}_{\boldsymbol{\beta}}$ and $U_{\theta}$ and $KI_{\theta\theta}$ is the scalar variance of $U_{\theta}$, so
\begin{align}\label{var1stage}
\Var(\hat{\theta})=
\frac{1}{I_{\theta\theta}}+\frac{\mathbf{I}_{\theta\boldsymbol{\beta}}(\mathbf{I}^{-1})_{\boldsymbol{\beta}\boldsymbol{\beta}}\mathbf{I}_{\boldsymbol{\beta}\theta}}{I_{\theta\theta}^2}.
\end{align}
In practical applications, standard errors of parameter estimates can be retrieved from the diagonal elements of the inverse of the Hessian matrix $\mathbf{I}$.
\subsection{Two-stage parametric estimation}
Two-stage parametric estimation, also referred to as the method of inference functions for margins \citep{Xu1996}, has been used mainly for multivariate models whenever a multi-parameter numerical optimization for maximum likelihood estimation is too time-consuming or infeasible. In the first stage, $\boldsymbol{\beta}$ is estimated by $\overline{\boldsymbol{\beta}}$ by considering all subjects as independent, identically distributed random variables, i.e. solving
$$\mathbf{U}^*_{\boldsymbol{\beta}}(\boldsymbol{\beta})=\sum_{i=1}^K \sum_{j=1}^{n_i} \delta_{ij} \frac{\partial \log f(x_{ij}|\boldsymbol{Z}_{ij})}{\partial \boldsymbol{\beta}} + (1-\delta_{ij}) \frac{\partial \log S(x_{ij}|\boldsymbol{Z}_{ij})}{\partial \boldsymbol{\beta}} =\boldsymbol{0}.$$
Under regularity conditions, $\sqrt{K}(\overline{\boldsymbol{\beta}}-\boldsymbol{\beta})$ converges to a multivariate normal distribution with mean vector zero and variance-covariance matrix $(\mathbf{I}^*)^{-1}\mathbf{V}(\mathbf{I}^*)^{-1}$, where $\mathbf{V}$ is the variance-covariance matrix of the score functions $\mathbf{U}^*_{\boldsymbol{\beta}}$ and $\mathbf{I}^*$ is the Fisher information of $\mathbf{U}^*_{\boldsymbol{\beta}}$. The use of the robust sandwich estimator is required since $(\mathbf{I}^*)^{-1}$ is not a consistent estimator of the asymptotic variance-covariance matrix due to the correlation between survival times. In the second stage, the association parameter $\theta$ is estimated by plugging in the estimates for the margins into the likelihood expression (\ref{eq3}), which is then maximized for the association parameter $\theta$. The two-stage estimator for $\theta$ is the solution to
$$U_{\theta}(\overline{\boldsymbol{\beta}},\theta)=\frac{\partial \log L}{\partial \theta}(\overline{\boldsymbol{\beta}},\theta)=0.$$
\begin{theorem}
Let $\overline{\theta}$ denote the solution to $U_{\theta}(\overline{\boldsymbol{\beta}},\theta)=0$ and let $\theta_0$ be the true value of the association parameter. Under regularity conditions, $\sqrt{K}(\overline{\theta}-\theta_0)$ converges to a normal distribution with mean zero and variance
\begin{align}\label{var2stage}
\Var(\overline{\theta}) = \frac{1}{I_{\theta \theta}}+ \frac{\mathbf{I}_{\theta\mathbf{\beta}}(\mathbf{I}^*)^{-1}\mathbf{V}(\mathbf{I}^*)^{-1} \mathbf{I}_{\mathbf{\beta}\theta}}{I_{\theta\theta}^2}.
\end{align}
\end{theorem}
\noindent To estimate this quantity, we make use of $(\mathbf{I}^*)^{-1}\mathbf{V}(\mathbf{I}^*)^{-1}$, the robust variance obtained in the first step, $I_{\theta\theta}^{-1}$ and $I_{\beta\theta}$ are obtained from the Hessian of the one-stage procedure. The proof of Theorem 1 is provided in the Appendix.
\subsection{Two-stage semiparametric estimation}
In the two-stage semiparametric estimation procedure, the marginal survival functions are estimated using the Cox proportional hazards model \citep{Cox1972}. Formulas for the standard error of the estimated covariate effect $\check{\boldsymbol{\beta}}$ and the estimated cumulative hazard $\check{\Lambda}$ that account for clustering can be found using a sandwich formula \citep{SpiekermanLin1998}.\\\\
In the second stage, $\max_{\theta}L(\theta;\check{\boldsymbol{\beta}},\check{\Lambda})$ is solved for $\check{\theta}$.
\begin{theorem}
Under regularity conditions C.1-C.7 in the Appendix, $(\check{\theta};\check{\boldsymbol{\beta}},\check{\Lambda})$ is a consistent estimator for $(\theta_0;\boldsymbol{\beta}_0,\Lambda_0)$.
 \end{theorem}
The results for $\check{\boldsymbol{\beta}}$ and $\check{\Lambda}$ follow from arguments along the lines of \citet{SpiekermanLin1998}. The consistency of $\check{\theta}$ is proved in the Appendix. Also following Spiekerman and Lin, one can show that $\sqrt{K}(\check{\boldsymbol{\beta}}-\boldsymbol{\beta}_0)$ converges to a mean zero normal distribution and that $\sqrt{K}(\check{\Lambda}-\Lambda_0)$ converges to a mean zero Gaussian process.
\begin{theorem}
Under regularity conditions C.1-C.7 in the Appendix, $\sqrt{K}(\check{\theta}-\theta_0)$ converges to a normal distribution with mean zero and variance
$$\frac{\Var(\Xi_1)}{W(\theta_0)^2}.$$
\end{theorem}
The proof of this theorem and the precise definition of $\Xi_1$ and $W(\theta_0)$, together with their estimators, can be found in the Appendix.
\section{Copula likelihood expression for distributions from the PVF family}\label{distributions}
The power variance function family of distributions, denoted PVF($\alpha, \delta, \gamma$), is a large class of distributions for which \citet{Hougaard2000} states that the Laplace transforms correspond to
$$\mathcal{L}(s)=\exp\left[-\frac{\delta}{\alpha}\left((\gamma + s)^\alpha - \gamma^\alpha\right)\right]$$
with derivatives
$$\mathcal{L}^{(k)}(s)=(-1)^k\mathcal{L}(s)\sum_{j=1}^k c_{k,j}(\alpha)\delta^j(\gamma+s)^{j\alpha - k},$$
where the coefficients $c_{k,j}(\alpha)$ are polynomials of order $k-j$ in $\alpha$, given by the recursive formula
$$c_{k,1}(\alpha)=\frac{\Gamma(k-\alpha)}{\Gamma(1-\alpha)}, \quad c_{k,k}=1$$
$$c_{k,j}(\alpha)=c_{k-1,j-1}(\alpha)+c_{k-1,j}(\alpha)(k-1-j\alpha)$$
This allows for a closed form expression of the copula likelihood (\ref{eq3}).\\
The one-parameter gamma distribution with density
$$g_{\theta}(x) = \frac{x^{1/\theta-1}e^{-x/\theta}}{\theta^{1/\theta}\Gamma(1/\theta)},\quad \theta>0.$$
is found as the limiting case $\alpha=0, \delta=\gamma=1/\theta$.
Failure times are independent when $\theta$ approaches zero. The Laplace transform is
$$\mathcal{L}(s) = \varphi_{\theta}(s)= (1+\theta s)^{-1/\theta}$$
which is the generator of the Clayton copula.\\
The choice $\alpha=\theta, \delta= \theta, \gamma=0$ leads to the positive stable distribution with density
$$g_{\theta}(x) = -\frac{1}{\pi x} \sum_{k=1}^{\infty} \frac{\Gamma(k\theta + 1)}{k!}(-x^{-\theta})^{k}\sin (\theta k \pi)$$
with $0 < \theta < 1$. \citet{Feller1971} shows that this density function can be found by Fourier inversion of the Laplace transform
$$\mathcal{L}(s) = \varphi_{\theta}(s)= e^{-s^{\theta}}$$
which is the generator of the Gumbel-Hougaard copula. Small values of $\theta$ provide large correlation and survival times are independent as $\theta$ approaches 1.\\
Another PVF distribution is obtained by choosing $\alpha=1/2, \delta=(2\theta)^{-1/2}, \gamma=(2\theta)^{-1}$. This is the inverse Gaussian distribution with variance $\theta$. The density is defined by
$$f_{\theta}(x) = \sqrt{\frac{1}{2\pi\theta}} x^{-3/2} \exp\left(\frac{-1}{2x\theta}(x-1)^2\right)$$
with $\theta >0$.
The Laplace transform is
$$\mathcal{L}(s) = \varphi_{\theta}(s)= \exp\left(\frac{1}{\theta}-\left(\frac{1}{\theta^2}+2\frac{s}{\theta}\right)^{1/2}\right).$$

\section{Modelling time to first insemination in cows clustered in herds}\label{dataexample}
In dairy cattle, the calving interval (the time between two calvings) should be optimally between 12 and 13 months. One of the main factors determining the length of the calving interval is the time from parturition to the time of first insemination \citep{DuchateauJanssen2004}. The objective of this study, amongst others, was to quantify the correlation between insemination times of cows within a herd. The data set includes 181 clusters (farms) of different sizes, ranging from $1$ cow to $174$ cows. The parity of the cow (0 if multiparous, 1 if primiparous) is known to be important, and is therefore added as a covariate. In the parametric approach, we assume a Weibull distribution for the times to first insemination $$S(t)=\exp(-\lambda\exp(\beta'Z)t^\rho)$$ and model the association structure by a Clayton copula.

The one-stage and two-stage parametric approach lead to similar results for the parity effect with hazard ratios equal to $0.92 \ (95\% \text{CI: }[0.89,0.95])$ and $0.94 \ (95\% \text{CI: }[0.90,0.98])$ respectively. The semiparametric two-stage approach leads to a hazard ratio of $0.94 \ (95\% \text{CI: }[0.90,0.98])$, the same as the one from the parametric two-stage approach. The parameter estimates for $\theta$ differ between the different approaches, with the lowest value observed for the one-stage parametric model and the highest, about the double, for the two-stage semiparametric model. Standard errors of one-stage parametric estimators are calculated from the inverse Hessian matrix. In the two-stage parametric approach, standard errors are found via formula (\ref{var2stage}). In the two-stage semiparametric case, we used the grouped jackknife to obtain standard errors \citep{LipsitzDearZao1994, LipsitzParzen1996}.
\begin{table}[h]
\footnotesize
\begin{center}
\begin{tabular}{c|c c c}
& Parametric & Parametric & Semiparametric \\
& one-stage & two-stage & two-stage\\
\hline
$\lambda$ & $0.00088(6.8\times 10^{-5})$&$0.00154(2.1\times 10^{-4})$\\
$\rho$ & $1.470(0.014)$&$1.344(0.033)$& \\
$\beta$ & $-0.082 (0.017)$&$-0.066(0.022)$&$-0.060(0.021)$\\
$\theta$ & $0.213(0.015)$&$0.324(0.052)$&$0.447(0.063)$
\end{tabular}
\caption{Estimation results for time to first insemination data}
\end{center}
\end{table}

\section{Simulation study}\label{simulation}
We generate 100 data sets with 50 or 200 clusters of size varying uniformly between 2 and 50. Survival times are simulated from a Clayton copula model with $\theta=0.2,0.5,1.0,1.5$ and from a Gumbel-Hougaard copula with $\theta=0.2,0.5,0.8$ and with Weibull marginal survival functions $S(t)=\lambda t^\rho \exp(\beta'Z)$, choosing $\rho=1.5$, $\lambda=0.0316$ and $Z$ a dichotomous covariate with effect $\beta=3$. Data are generated using the sampling algorithm of \citet{MarshallOlkin1988}. The censoring distribution is also Weibull, with parameters ($\lambda_C=0.0274, \rho_C=1.5)$ and $(\lambda_C=0.1464, \rho_C=1.5)$ yielding censoring percentages of $25\%$ and $50\%$, respectively. The performances of one-stage parametric estimation, two-stage parametric estimation and two-stage semi-parametric estimation are summarized in Tables 2 and 3. For each copula and value of $\theta_0$, we report the mean value of $\hat{\theta}$, $\overline{\theta}$ and $\check{\theta}$ in the first row. Mean standard errors together with the coverage are reported in the second row. As the number of clusters increases from $K=50$ to $K=200$, standard errors are halved since they are proportional to $1/\sqrt{K}$. The estimates are not noticeably affected by an increasing percentage of censoring, instead we observe that biases tend to shrink as $\theta_0$ approaches independence. The largest biases are found in the semiparametric cases where $\theta_0$ has moved far away from independence. The transition from $K=50$ to $K=200$ leads to a reduction of the bias, which also follows from the asymptotic proofs in the Appendix. Although computationally more demanding, the one-stage parametric procedure yields the best coverages in all cases except the Gumbel-Hougaard copula with $\theta_0=0.2$.

\section{Discussion}
The current copula methodology only allows the modelling of multivariate survival data that are grouped in clusters of small and equal size. A new formulation for the likelihood of Archimedean copula models for survival data is developed, that allows for clusters of large and variable size. The failure times within a cluster are assumed to be exchangeable and the whole data set is used to estimate a common marginal baseline survival. The survival functions of subjects differ through the incorporation of covariates (possibly time-dependent). For copula members of the PVF family, a closed form expression of the likelihood exists, whereas other choices require numerical integration.
We investigated the parametric one-stage and two-stage approach as well as the semiparametric two-stage approach and derived asymptotic results for the estimators under a reasonable set of conditions. Simulation results show that all three methods work well for cluster sizes ranging from 2 to 50. Even larger clusters can be attained, at the cost of larger computing time. This article is an extension of the work of \citet{ShihLouis1995}, who derived founding results for bivariate data, and the work of \citet{Glidden2000}, who investigated the two-stage semiparametric model for the Clayton copula, as it describes the use of copula functions for clusters with large and varying cluster size.
\begin{landscape}
\begin{table}[H]
\scriptsize
\begin{tabular}{c|c|c c c|c c c|c c c}
& &  \multicolumn{3}{c}{$0\%$ censoring} & \multicolumn{3}{c}{$25\%$ censoring}  & \multicolumn{3}{c}{$50\%$ censoring} \\
Copula &  & Parametric & Parametric & Semiparametric & Parametric & Parametric & Semiparametric & Parametric & Parametric & Semiparametric \\
model& $\theta_0$ & one-stage & two-stage & two-stage & one-stage & two-stage & two-stage & one-stage & two-stage & two-stage\\
\hline
Clayton&$0.2$&$0.206$&$0.203$&$0.201$&$0.207$&$0.206$&$0.204$&$0.207$&$0.206$&$0.207$\\
&&$(0.043;96\%)$&$(0.044;94\%)$&$(0.047;90\%)$&$(0.049;96\%)$&$(0.049;96\%)$&$(0.049;96\%)$&$(0.056;98\%)$&$(0.057;98\%)$&
$(0.060,93\%)$\\
&$0.5$&$0.501$&$0.495$&$0.472$&$0.501$&$0.496$&$0.490$&$0.506$&$0.502$&$0.499$\\
&&$(0.084;96\%)$&$(0.093;87\%)$&$(0.099;79\%)$&$(0.091;96\%)$&$(0.099;92\%)$&$(0.104;84\%)$&$(0.102;95\%)$&$(0.108;93\%)$&$(0.114;93\%)$\\
&$1.0$&$1.016$&$0.984$&$0.873$&$1.021$&
$0.994$&$0.948$&$1.015$&$0.995$&$0.962$\\
&&$(0.162;94\%)$&$(0.168;86\%)$&$(0.160;72\%)$&$(0.170;93\%)$&$(0.174;86\%)$&$(0.185;83\%)$&$(0.180;94\%)$&$(0.185;91\%)$&$(0.194;85\%)$\\
&$1.5$&$1.476$&$1.429$&$1.205$&$1.475$&$1.450$&$1.351$&$1.475$&$1.473$&$1.385$\\
&&$(0.235;90\%)$&$(0.254;81\%)$&$(0.223;54\%)$&$(0.240;92\%)$&$(0.261;87\%)$&$(0.269;78\%)$&$(0.252;91\%)$&$(0.279;87\%)$&$(0.281;83\%)$\\
\hline
G-H&$0.2$&$0.193$&$0.203$&$0.245$&$0.203$&$0.205$&$0.247$&$0.202$&$0.207$&$0.256$\\
&&$(0.020;87\%)$&$(0.024;97\%)$&$(0.029;68\%)$&$(0.011;100\%)$&$(0.024;95\%)$&$(0.015;67\%)$&$(0.022;97\%)$&$(0.025;94\%)$&$(0.035;62\%)$\\
&$0.5$&$0.505$&$0.503$&$0.516$&$0.507$&$0.503$&$0.518$&$0.506$&$0.504$&$0.521$\\
&&$(0.040;94\%)$&$(0.046;96\%)$&$(0.047;94\%)$&$(0.041;95\%)$&$(0.049;93\%)$&$(0.051;94\%)$&$(0.043;96\%)$&$(0.052;92\%)$&$(0.054;90\%)$\\
&$0.8$&$0.805$&$0.799$&$0.801$&$0.805$&$0.799$&$0.800$&$0.805$&$0.800$&$0.801$\\ &&$(0.034;92\%)$&$(0.043;91\%)$&$(0.043;91\%)$&$(0.036;94\%)$&$(0.046;91\%)$&$(0.046;90\%)$&$(0.039;95\%)$&$(0.049;90\%)$&$(0.050;89\%)$
\end{tabular}
\caption{Simulation results for $K=50$}\label{sim50}
\end{table}
\begin{table}[H]
\scriptsize
\begin{tabular}{c|c|c c c|c c c|c c c}
& &  \multicolumn{3}{c}{$0\%$ censoring} & \multicolumn{3}{c}{$25\%$ censoring}  & \multicolumn{3}{c}{$50\%$ censoring} \\
Copula &  & Parametric & Parametric & Semiparametric & Parametric & Parametric & Semiparametric & Parametric & Parametric & Semiparametric \\
model& $\theta_0$ & one-stage & two-stage & two-stage & one-stage & two-stage & two-stage & one-stage & two-stage & two-stage\\
\hline
Clayton&$0.2$&$0.200$&$0.198$&$0.196$&$0.199$&$0.197$&$0.197$&$0.200$&$0.199$&$0.198$\\
&&$(0.021;97\%)$&$(0.022;96\%)$&$(0.024;90\%)$&$(0.024;97\%)$&$(0.024;97\%)$&$(0.025;96\%)$&$(0.027;99\%)$&$(0.027;96\%)$&$(0.028;94\%)$\\
&$0.5$&$0.498$&$0.496$&$0.487$&$0.497$&$0.495$&$0.492$&$0.494$&$0.492$&$0.490$\\
&&$(0.042;96\%)$&$(0.050;93\%)$&$(0.056;87\%)$&$(0.045;94\%)$&$(0.050;94\%)$&$(0.055;91\%)$&$(0.050;92\%)$&$(0.053;94\%)$&$(0.057;90\%)$\\
&$1.0$&$1.002$&$0.996$&$0.956$&$0.997$&$0.993$&$0.981$&$0.998$&$0.995$&$0.986$\\
&&$(0.080;95\%)$&$(0.100;92\%)$&$(0.108;85\%)$&$(0.083;93\%)$&$(0.099;94\%)$&$(0.106;90\%)$&$(0.089;95\%)$&$(0.101;94\%)$&$(0.108;92\%)$\\
&$1.5$&$1.482$&$1.488$&$1.408$&$1.482$&$1.490$&$1.468$&$1.491$&$1.496$&$1.481$\\
&&$(0.117;94\%)$&$(0.150;88\%)$&$(0.154;82\%)$&$(0.120;95\%)$&$(0.146;88\%)$&$(0.157;86\%)$&$(0.127;95\%)$&$(0.149;89\%)$&$(0.159;89\%)$\\
\hline
G-H&$0.2$&$0.195$&$0.203$&$0.218$&$0.202$&$0.204$&$0.219$&$0.203$&$0.204$&$0.222$\\
&&$(0.011;84\%)$&$(0.012;93\%)$&$(0.014;77\%)$&$(0.011;97\%)$&$(0.013;95\%)$&$(0.031;80\%)$&$(0.011;97\%)$&$(0.014;91\%)$&$(0.016;77\%)$\\
&$0.5$&$0.504$&$0.503$&$0.508$&$0.503$&$0.502$&$0.507$&$0.503$&$0.502$&$0.507$\\
&&$(0.020;96\%)$&$(0.024;93\%)$&$(0.024;94\%)$&$(0.020;95\%)$&$(0.026;93\%)$&$(0.026;95\%)$&$(0.022;98\%)$&$(0.028;93\%)$&$(0.029;94\%)$\\ &$0.8$&$0.802$&$0.799$&$0.799$&$0.802$&$0.798$&$0.798$&$0.801$&$0.797$&$0.797$\\
&&$(0.017;92\%)$&$(0.023;91\%)$&$(0.023;92\%)$&$(0.018;93\%)$&$(0.025;92\%)$&$(0.025;92\%)$&$(0.020;96\%)$&$(0.027;93\%)$&$(0.027;92\%)$
\end{tabular}
\caption{Simulation results for $K=200$}\label{sim200}
\end{table}
\end{landscape}
\bibliographystyle{apalike}
\bibliography{UnbalancedCopula}

\begin{thebibliography}{}

\bibitem[Andersen, 2005]{Andersen2005}
Andersen, E.~W. (2005).
\newblock Two-stage estimation in copula models used in family studies.
\newblock {\em Lifetime Data Analysis}, 11:333--350.

\bibitem[Cox, 1972]{Cox1972}
Cox, D.~R. (1972).
\newblock Regression models and life-tables.
\newblock {\em Journal of the Royal Statistical Society}, 34:187--220.

\bibitem[Cox and Hinkley, 1974]{CoxHinkley1974}
Cox, D.~R. and Hinkley, D. (1974).
\newblock {\em Theoretical Statistics}.
\newblock Chapman and Hall.

\bibitem[Duchateau and Janssen, 2004]{DuchateauJanssen2004}
Duchateau, L. and Janssen, P. (2004).
\newblock Penalized partial likelihood for frailties and smoothing splines in
  time to first insemination models for dairy cows.
\newblock {\em Biometrics}, 60(3):608--614.

\bibitem[Duchateau and Janssen, 2008]{DuchateauJanssen2008}
Duchateau, L. and Janssen, P. (2008).
\newblock {\em The Frailty Model}.
\newblock Springer.

\bibitem[Feller, 1971]{Feller1971}
Feller, W. (1971).
\newblock {\em An Introduction to Probability Theory and Its Applications}.
\newblock Wiley.

\bibitem[Glidden, 2000]{Glidden2000}
Glidden, D.~V. (2000).
\newblock A two-stage estimator of the dependence parameter for the
  clayton-oakes model.
\newblock {\em Lifetime Data Analysis}, 6:141--156.

\bibitem[Hougaard, 2000]{Hougaard2000}
Hougaard, P. (2000).
\newblock {\em Analysis of Multivariate Survival Data}.
\newblock Springer.

\bibitem[Joe, 1997]{Joe1997}
Joe, H. (1997).
\newblock {\em Multivariate Models and Dependence Concepts}.
\newblock Chapman \& Hall.

\bibitem[Joe, 2005]{Joe2005}
Joe, H. (2005).
\newblock Asymptotic efficiency of the two-stage estimation method for
  copula-based models.
\newblock {\em Journal of Multivariate Analysis}, 94:401--419.

\bibitem[Lipsitz et~al., 1994]{LipsitzDearZao1994}
Lipsitz, S.~R., Dear, K.~B., and Zhao, L. (1994).
\newblock Jackknife estimators of variance for parameter estimates from
  estimating equations with applications to clustered survival data.
\newblock {\em Biometrics}, 50:842--846.

\bibitem[Lipsitz and Parzen, 1996]{LipsitzParzen1996}
Lipsitz, S.~R. and Parzen, M. (1996).
\newblock A jackknife estimator of variance for cox regression for correlated
  survival data.
\newblock {\em Biometrics}, 52:291--298.

\bibitem[Marshall and Olkin, 1988]{MarshallOlkin1988}
Marshall, A.~W. and Olkin, I. (1988).
\newblock Families of multivariate distributions.
\newblock {\em Journal of the American Statistical Association}, 83(403):834 --
  841.

\bibitem[Massonnet et~al., 2009]{MassonnetJanssenDuchateau2009}
Massonnet, G., Janssen, P., and Duchateau, L. (2009).
\newblock Modelling udder infection data using copula models for quadruples.
\newblock {\em Journal of Statistical Planning and Inference}, 139:3865 --3877.

\bibitem[Nelsen, 2006]{Nelsen2006}
Nelsen, R.~B. (2006).
\newblock {\em An Introduction to copulas}.
\newblock Springer.

\bibitem[Othus and Li, 2010]{OthusLi2010}
Othus, M. and Li, Y. (2010).
\newblock A gaussian copula model for multivariate survival data.
\newblock {\em Statistics in Biosciences}, 2:154--179.

\bibitem[Shih and Louis, 1995]{ShihLouis1995}
Shih, J.~H. and Louis, T.~A. (1995).
\newblock Inferences on the association parameter in copula models for
  bivariate survival data.
\newblock {\em Biometrics}, 51:1384--1399.

\bibitem[Spiekerman and Lin, 1998]{SpiekermanLin1998}
Spiekerman, C.~F. and Lin, D.~Y. (1998).
\newblock Marginal regression models for multivariate failure time data.
\newblock {\em Journal of the American Statistical Association}, 93:1164--1175.

\bibitem[Wienke, 2011]{Wienke2011}
Wienke, A. (2011).
\newblock {\em Frailty Models in Survival Analysis}.
\newblock Chapman \& Hall.

\bibitem[Xu, 1996]{Xu1996}
Xu, J. (1996).
\newblock Statistical modelling and inference for multivariate and longitudinal
  discrete response data.
\newblock {\em Ph.D. Thesis}.

\end{thebibliography}

\section*{Appendix: Theorems and proofs}
\footnotesize
\textbf{Proof of Theorem 1.} Let $\boldsymbol{\beta}_0$ denote the true parameter vector for the margins. Expanding the score function $\mathbf{U}^*_{\boldsymbol{\beta}}$ in a Taylor series around $\boldsymbol{\beta}_0$ and evaluating it at $\boldsymbol{\beta}=\overline{\boldsymbol{\beta}}$, we get under regularity conditions of maximum likelihood theory
$$\mathbf{U}^*_{\boldsymbol{\beta}}(\overline{\boldsymbol{\beta}}) = \mathbf{0} = \mathbf{U}^*_{\boldsymbol{\beta}}(\boldsymbol{\beta}_0) + \left.\frac{\partial \mathbf{U}^*_{\boldsymbol{\beta}}}{\partial \boldsymbol{\beta}}\right|_{\boldsymbol{\beta}=\boldsymbol{\beta}_0}(\overline{\boldsymbol{\beta}}-\boldsymbol{\beta}_0) + o_p(\sqrt{K}).$$
Similarly,
$$U_{\theta}(\overline{\boldsymbol{\beta}},\overline{\theta})= 0 = U_{\theta}(\boldsymbol{\beta}_0,\theta_0) + \left.\frac{\partial U_\theta}{\partial \boldsymbol{\beta}}\right|_{(\boldsymbol{\beta},\theta)=(\boldsymbol{\beta}_0,\theta_0)}(\overline{\boldsymbol{\beta}}-\boldsymbol{\beta}_0)
+ \left.\frac{\partial U_\theta}{\partial \theta}\right|_{(\boldsymbol{\beta},\theta)=(\boldsymbol{\beta}_0,\theta_0)}(\overline{\theta}-\theta_0) + o_p(\sqrt{K}).$$
By the law of large numbers, as $K \to \infty$,
\begin{align*}
& -\frac{1}{K}\left.\frac{\partial \mathbf{U}^*_{\boldsymbol{\beta}}}{\partial \boldsymbol{\beta}}\right|_{\boldsymbol{\beta}=\boldsymbol{\beta}_0}=\frac{1}{K}\sum_{i=1}^K - \frac{\partial}{\partial \boldsymbol{\beta}}\mathbf{U}^*_{i,\boldsymbol{\beta}}(\boldsymbol{\beta}_0) \to \mathbf{I}^*=E\left[- \frac{\partial}{\partial \boldsymbol{\beta}}\mathbf{U}^*_{1,\boldsymbol{\beta}}(\boldsymbol{\beta}_0)\right]
\\
& -\frac{1}{K}\left.\frac{\partial U_\theta}{\partial \boldsymbol{\beta}}\right|_{(\boldsymbol{\beta},\theta)=(\boldsymbol{\beta}_0,\theta_0)} = \frac{1}{K}\sum_{i=1}^K-\frac{\partial}{\partial{\boldsymbol{\beta}}}U_{i,\theta}(\boldsymbol{\beta}_0,\theta_0)
\to \mathbf{I}_{\theta\boldsymbol{\beta}}\\
& -\frac{1}{K} \left.\frac{\partial U_\theta}{\partial \theta}\right|_{(\boldsymbol{\beta},\theta)=(\boldsymbol{\beta}_0,\theta_0)} = \frac{1}{K}\sum_{i=1}^K-\frac{\partial}{\partial{\theta}}U_{i,\theta}(\boldsymbol{\beta}_0,\theta_0)\to I_{\theta\theta}.
\end{align*}
Hence $$\frac{1}{\sqrt{K}}\left(\begin{array}{c}\mathbf{U}^*_{\boldsymbol{\beta}}(\boldsymbol{\beta}_0)\\
U_{\theta}(\boldsymbol{\beta}_0,\theta_0)\end{array}\right) \to \sqrt{K}\left(\begin{array}{c c} \mathbf{I}^* & 0\\
\mathbf{I}_{\theta\boldsymbol{\beta}} & I_{\theta\theta} \end{array}\right) \left(\begin{array}{c}\overline{\boldsymbol{\beta}}-\boldsymbol{\beta}_0\\
\overline{\theta}-\theta_0\end{array}\right).$$
By the central limit theorem, $\frac{1}{\sqrt{K}}\left(\begin{array}{c}\mathbf{U}^*_{\boldsymbol{\beta}}(\boldsymbol{\beta}_0)\\
U_{\theta}(\boldsymbol{\beta}_0,\theta_0)\end{array}\right)$ converges to multivariate normal with mean $\left(\begin{array}{c}0\\0 \end{array}\right)$
and variance-covariance matrix
$\left(\begin{array}{c c}
\mathbf{V} & 0\\
0 & I_{\theta\theta}
\end{array}\right)$
with $\mathbf{V}=\Var\left(\mathbf{U}^*_{1,\boldsymbol{\beta}}(\boldsymbol{\beta}_0)\right)
=E\left[\mathbf{U}^*_{1,\boldsymbol{\beta}}(\boldsymbol{\beta}_0)^2\right]$.
Thus, $\sqrt{K} \left(\begin{array}{c}\overline{\boldsymbol{\beta}}-\boldsymbol{\beta}_0\\
\overline{\theta}-\theta_0\end{array}\right)$ converges to multivariate normal with mean vector zero and variance-covariance matrix
\begin{align*}
&\left(\begin{array}{c c}
\mathbf{I}^* & \mathbf{0}\\
\mathbf{I}_{\theta\boldsymbol{\beta}} & I_{\theta\theta}
\end{array}\right)^{-1}
\left(\begin{array}{c c}
\mathbf{V} & \mathbf{0}\\
\mathbf{0} & I_{\theta\theta}
\end{array}\right)
{\left(\begin{array}{c c}
\mathbf{I}^* & \mathbf{0}\\
\mathbf{I}_{\theta\boldsymbol{\beta}} & I_{\theta\theta}
\end{array}\right)^{-1}}^T=\left(\begin{array}{c c}
(\mathbf{I}^*)^{-1}\mathbf{V}{(\mathbf{I}^*)^{-1}}^T & \frac{-{(\mathbf{I}^*)^{-1}\mathbf{V}(\mathbf{I}^*)^{-1}}^T \mathbf{I}_{\boldsymbol{\beta}\theta}}{I_{\theta\theta}}\\
\frac{-\mathbf{I}_{\theta\boldsymbol{\beta}}{(\mathbf{I}^*)^{-1}\mathbf{V}(\mathbf{I}^*)^{-1}}^T}{I_{\theta\theta}} & \frac{1}{I_{\theta\theta}}+\frac{\mathbf{I}_{\theta\boldsymbol{\beta}}(\mathbf{I}^*)^{-1}\mathbf{V}{(\mathbf{I}^*)^{-1}}^T \mathbf{I}_{\boldsymbol{\beta}\theta}}{I_{\theta\theta}^2}
\end{array}\right).
\end{align*}
The lower right element of this matrix is the asymptotic variance of $\sqrt{K}(\overline{\theta}-\theta_0)$ and we denote this by $\sigma^2$.
$$\sigma^2 = \frac{1}{I_{\theta\theta}}+ \frac{\mathbf{I}_{\theta\boldsymbol{\beta}}(\mathbf{I}^*)^{-1}\mathbf{V}(\mathbf{I}^*)^{-1} \mathbf{I}_{\boldsymbol{\beta}\theta}}{I_{\theta\theta}^2}.$$
Before we prove Theorem 2 and 3, we first introduce some notation.
\begin{align*}
Y_{ij}(t)&=I_{\{X_{ij}\geq t\}} \\
\check{\Lambda}(t)&=\int_0^t \frac{d\sum_{i=1}^K \sum_{j=1}^{n_i}\delta_{ij}I_{\{X_{ij}\leq u\}}}{\sum_{i=1}^K\sum_{j=1}^{n_i}Y_{ij}(u)\exp[\check{\boldsymbol{\beta}}'\boldsymbol{Z}_{ij}(u)]} =\sum_{i=1}^K\sum_{j=1}^{n_i}\frac{\delta_{ij}I_{\{X_{ij}\leq t\}}}{\sum_{k=1}^K\sum_{l=1}^{n_k}I_{\{X_{kl}\leq X_{ij}\}}\exp[\check{\boldsymbol{\beta}}'\boldsymbol{Z}_{kl}(X_{ij})]}\\
H_{ij}&=\exp\left(-\int_0^\tau Y_{ij}(u) \exp[\boldsymbol{\beta}'\boldsymbol{Z}_{ij}(u)]d\Lambda(u)\right)\\
H_{ij}^0&=\exp\left(-\int_0^\tau Y_{ij}(u) \exp[\boldsymbol{\beta}_0'\boldsymbol{Z}_{ij}(u)]d\Lambda_0(u)\right)\\
\check{H}_{ij}&=\exp\left(-\int_0^\tau Y_{ij}(u) \exp[\check{\boldsymbol{\beta}}'\boldsymbol{Z}_{ij}(u)]d\check{\Lambda}(u)\right)\\
H_{ij}(t) &=\exp\left(-\int_0^{\tau}Y_{ij}(u)\exp[\boldsymbol{\beta}'\boldsymbol{Z}_{ij}(u)]d(\Lambda+t(\Gamma-\Lambda))(u)\right) \end{align*}
Note that $H_{ij}=H_{ij}(0)$.
\begin{align*}
L(\theta;\boldsymbol{\beta},\Lambda) &= \prod_{i=1}^K L_i(\theta;\boldsymbol{\beta},\Lambda)\\
&=\prod_{i=1}^K\left(\prod_{j=1}^{n_i}\left[\frac{1}{\varphi_{\theta}'\left(\varphi_{\theta}^{-1}\left(H_{ij}
\right)\right)}\right]^{\delta_{ij}}\right)
\varphi_{\theta}^{(d_i)}\left(\sum_{j=1}^{n_i}\varphi_{\theta}^{-1}\left(H_{ij}\right)\right)\\
l_K(\theta)&=K^{-1}\log L(\theta;\boldsymbol{\beta},\Lambda)\\
&=K^{-1}\sum_{i=1}^{K}\left\{\sum_{j=1}^{n_i} \delta_{ij} \log \left[\frac{1}{\varphi_{\theta}'\left(\varphi_{\theta}^{-1}\left(H_{ij}\right)\right)} \right]+\log \varphi_{\theta}^{(d_i)}\left(\sum_{j=1}^{n_i} \varphi_{\theta}^{-1}\left(H_{ij}\right)\right)\right\}\\
l_{K0}(\theta)&=K^{-1}\log L(\theta;\boldsymbol{\beta}_0,\Lambda_0)\\
\check{l}_K(\theta)&=K^{-1}\log L(\theta;\check{\boldsymbol{\beta}},\check{\Lambda})\\
U_K(\theta)&=\frac{\partial l_K(\theta)}{\partial \theta}=K^{-1}\frac{\partial\log L(\theta;\boldsymbol{\beta},\Lambda)}{\partial \theta}\\
&=K^{-1}\sum_{i=1}^{K}\left\{\sum_{j=1}^{n_i} \delta_{ij}  \left[\varphi_{\theta}'\left(\varphi_{\theta}^{-1}\left(H_{ij}\right)\right) \right] \frac{\partial}{\partial \theta}\left[\varphi_{\theta}'\left(\varphi_{\theta}^{-1}\left(H_{ij}\right)\right) \right]^{-1}\right.\\
&\qquad \left.+\left[\varphi_{\theta}^{(d_i)}\left(\sum_{j=1}^{n_i} \varphi_{\theta}^{-1}\left(H_{ij}\right)\right)\right]^{-1} \frac{\partial}{\partial \theta}\left[\varphi_{\theta}^{(d_i)}\left(\sum_{j=1}^{n_i} \varphi_{\theta}^{-1}\left(H_{ij}\right)\right)\right]\right\}\\
U_{K0}(\theta)&=\frac{\partial l_{K0}(\theta)}{\partial \theta}=K^{-1}\frac{\partial\log L(\theta;\boldsymbol{\beta}_0,\Lambda_0)}{\partial \theta}\\
\check{U}_K(\theta)&=\frac{\partial \check{l}_K(\theta)}{\partial \theta}=K^{-1}\frac{\partial\log L(\theta;\check{\boldsymbol{\beta}},\check{\Lambda})}{\partial \theta}\\
\end{align*}
We copy the following notation from \citet{SpiekermanLin1998} where $\boldsymbol{a}^{\otimes 0}=1, \boldsymbol{a}^{\otimes 1}=\boldsymbol{a}$ and $\boldsymbol{a}^{\otimes 2}=\boldsymbol{a}'\boldsymbol{a}$:
\begin{align*}
&\boldsymbol{S}^{(r)}(\boldsymbol{\beta},t)=K^{-1}\sum\limits_{i=1}^K\sum\limits_{j=1}^{n_i}Y_{ij}(t)\exp[\boldsymbol{\beta}'\boldsymbol{Z}_{ij}(t)]\boldsymbol{Z}_{ij}(t)^{\otimes r}, \qquad 
\boldsymbol{s}^{(r)}=E\left[\boldsymbol{S}^{(r)}(\boldsymbol{\beta},t)\right] \quad (r=0,1,2)\\
&\boldsymbol{E}(\boldsymbol{\beta},t)
=\frac{\boldsymbol{S}^{(1)}(\boldsymbol{\beta},t)}{S^{(0)}(\boldsymbol{\beta},t)},\qquad  
\boldsymbol{e}(\boldsymbol{\beta},t)=\frac{\boldsymbol{s}^{(1)}(\boldsymbol{\beta},t)}{s^{(0)}(\boldsymbol{\beta},t)}\\
&\boldsymbol{V}(\boldsymbol{\beta},t)=\frac{\boldsymbol{S}^{(2)}(\boldsymbol{\beta},t)}{S^{(0)}(\boldsymbol{\beta},t)}-\boldsymbol{E}(\boldsymbol{\beta},t)^{\otimes 2}, \qquad 
\boldsymbol{v}(\boldsymbol{\beta},t)=\frac{\boldsymbol{s}^{(2)}(\boldsymbol{\beta},t)}{s^{(0)}(\boldsymbol{\beta},t)}-\boldsymbol{e}(\boldsymbol{\beta},t)^{\otimes 2}
\end{align*}
Assume the following regularity conditions where $\tau>0$ is a constant (e.g. end of study time).
 \begin{enumerate}
 \item[C1.] $\boldsymbol{\beta}$ is in a compact subset of $\mathbb{R}^p$
 \item[C2.] $\Lambda(\tau)<\infty$
 \item[C3.] $\theta \in \nu$, where $\nu$ is a compact subset of $\Theta$ \label{C2}
 \item[C4.] $P(C_{ij}\geq t \quad \forall t \in [0,\tau])>\delta_c>0$ for $i=1,\dots,K$ and $j=1,\dots,n_i$
 \item[C5.] Write $\boldsymbol{Z}_{ij}(t)=\{Z_{ij1}(t),\dots,Z_{ijp}(t)\}$. For $i=1,\dots,K,j=1\dots,n_i,k=1,\dots,p$ $$\left|Z_{ijk}(0)\right|+\int_0^\tau\left|dZ_{ijk}(t)\right|\leq B_Z<\infty \quad \text{a.s. for some constant }B_Z$$
 \item[C6.] $\displaystyle E\left[\log\frac{L_i(\theta_1;\boldsymbol{\beta},\Lambda)}{L_i(\theta_2;\boldsymbol{\beta},\Lambda)}\right]$ exists for all $\theta_1,\theta_2 \in \Theta, i=1,\dots,K$
 \item[C7.] $\boldsymbol{A}=\int_0^\tau \boldsymbol{v}(\boldsymbol{\beta}_0,u)s^{(0)}(\boldsymbol{\beta}_0,u)d\Lambda_0(u)$ is positive definite.
 \end{enumerate}
\textbf{Proof of Theorem 2.}
The results for $\check{\boldsymbol{\beta}}$ and $\check{\Lambda}$ follow from arguments along the lines of \citet{SpiekermanLin1998}. We will now show the consistency of $\check{\theta}$ using ideas of \citet{OthusLi2010}.\\\\
 To account for the fact that plug-in estimates of $\boldsymbol{\beta}$ and $\Lambda$ are used in the likelihood for $\theta$, we will need to take a Taylor series expansion of the likelihood of $\theta$ around $\boldsymbol{\beta}_0$ and $\Lambda_0$. Since $\Lambda_0$ is an unspecified function, this expansion will need to include a functional expansion term. An expansion using Hadamard derivatives is appropriate for this situation. Hereto, we must verify that the log-likelihood $l_K(\theta)$ is Hadamard differentiable with respect to $\Lambda$.\\\\
We find the Hadamard derivative of $l_{K}$ w.r.t. $\Lambda$ at $\Gamma-\Lambda \in BV[0,\tau]$ by taking the derivative of $K^{-1}\log L(\theta;\boldsymbol{\beta},\Lambda+t(\Gamma-\Lambda))$ with respect to $t$ en then putting $t=0$:
\begin{align*}
\left.\frac{d}{dt}\left[K^{-1}\log L(\theta;\boldsymbol{\beta},\Lambda+t(\Gamma-\Lambda))\right]\right|_{t=0}
=\int_0^\tau \zeta_K(\theta;\Lambda)(u)d(\Gamma-\Lambda)(u)
\end{align*}
where
\begin{align*}\zeta_K(\theta;\Lambda)(u) & =
K^{-1}\sum_{i=1}^K\sum_{j=1}^{n_i}D_{ij}^l Y_{ij}(u)\exp[\boldsymbol{\beta}'\boldsymbol{Z}_{ij}(u)]
\end{align*}
and
\begin{align*}
D_{ij}^l=\left\{\delta_{ij} \frac{-\varphi_\theta''(\varphi_\theta^{-1}(H_{ij}))}{\varphi_\theta'(\varphi_\theta^{-1}(H_{ij}))}
+ \frac{\varphi_\theta^{(d_i+1)}\left(\sum_{j=1}^{n_i}\varphi_\theta^{-1}(H_{ij})\right)}
{\varphi_\theta^{(d_i)}\left(\sum_{j=1}^{n_i}\varphi_\theta^{-1}(H_{ij})\right)}
\right\}\frac{-H_{ij}}{\varphi_\theta'(\varphi_\theta^{-1}(H_{ij}))}.
\end{align*}
The derivative of $l_K(\theta)$ w.r.t. $\boldsymbol{\beta}$ is
\begin{align*} \zeta_K(\theta;\boldsymbol{\beta}) =
K^{-1}\sum_{i=1}^K\sum_{j=1}^{n_i} D_{ij}^l \left(\int_0^\tau Y_{ij}(u)\boldsymbol{Z}_{ij}(u)\exp[\boldsymbol{\beta}'\boldsymbol{Z}_{ij}(u)]d\Lambda(u) \right).
\end{align*}
\noindent To prove consistency for $\check{\theta}$, we will require $||\zeta_K(\theta;\Lambda)||_\infty$ and $||\zeta_K(\theta;\boldsymbol{\beta})||$ to be bounded. This can be obtained when the common factor $||D_{ij}^l||_\infty$ is bounded and also the terms unique to $\zeta_K(\theta;\boldsymbol{\beta})$ and $\zeta_K(\theta;\Lambda)$ have to be bounded. This requirement is not too restrictive, e.g. for the Clayton copula we have
$$||D_{ij}^l||_\infty = \displaystyle\left|\left|\delta_{ij}(1+\theta)-\frac{(1+d_i\theta)H_{ij}^{-\theta}}{\left(-n_i+1+ \sum_{j=1}^{n_i}H_{ij}^{-\theta}\right)}\right|\right|_\infty.$$
Due to the definition of $H_{ij}$ and condition C2, this expression is bounded. By condition C5,
$$||Y_{ij}\exp[\boldsymbol{\beta}'\boldsymbol{Z}_{ij}]||_\infty \quad \text{and} \quad \left|\left|\int_0^\tau Y_{ij}(u)\boldsymbol{Z}_{ij}(u)\exp[\boldsymbol{\beta}'\boldsymbol{Z}_{ij(u)}d\Lambda(u)]\right|\right| \quad \text{are bounded.}$$
An expansion of $\check{l}_K(\theta)$ around $\boldsymbol{\beta}_0$ and $\Lambda_0$ can be written as
$$\check{l}_K(\theta)=l_{K0}(\theta)+\zeta_K(\theta;\boldsymbol{\beta}_0)(\check{\boldsymbol{\beta}}-\boldsymbol{\beta}_0)+\int_0^\tau \zeta_K(\theta;\Lambda_0)(t)d(\check{\Lambda}-\Lambda_0)(t) + R.$$
Another (intuitive) notation is:
$$l_{K,\theta}(\check{\boldsymbol{\beta}},\check{\Lambda})=l_{K,\theta}(\boldsymbol{\beta}_0,\Lambda_0)+\frac{\partial}{\partial \boldsymbol{\beta}} l_{K,\theta}(\boldsymbol{\beta}_0,\Lambda_0)(\check{\boldsymbol{\beta}}-\boldsymbol{\beta}_0)+ \frac{\partial}{\partial \Lambda} l_{K,\theta}(\boldsymbol{\beta}_0,\Lambda_0)(\check{\Lambda}-\Lambda_0) + R.$$
The remainder term $R$ is of order $o_p\left(\max\{||\check{\boldsymbol{\beta}}-\boldsymbol{\beta}_0||,||\check{\Lambda}-\Lambda_0||_\infty\} \right)$. This can be seen from the definition of Hadamard differentiability, since
$$\left|\left| \frac{l_{K,\theta}(\boldsymbol{\beta},\Lambda_0+t(\check{\Lambda}-\Lambda_0)) - l_{K,\theta}(\boldsymbol{\beta},\check{\Lambda})}{t} - \frac{\partial}{\partial \Lambda}l_{K,\theta}(\boldsymbol{\beta},\Lambda_0)(\check{\Lambda}-\Lambda_0) \right|\right|_{\infty} \to 0,\qquad \text{as }t\downarrow 0,$$
uniformly in $\check{\Lambda}-\Lambda_0$ in all compact subsets of $\mathbb{D}$, the space of cumulative hazard functions. Since $\check{\boldsymbol{\beta}}$ is consistent and $\check{\Lambda}$ is uniformly consistent \citep{SpiekermanLin1998}, $R=o_p(1)$.
\\\\
In order to prove $\check{\theta}$ is consistent we will need to verify the uniform convergence of the log-likelihood with the plug-in estimate of $\Lambda$ to the expected value of the log-likelihood evaluated at the true value of $\Lambda$, denoted $l_{K0}(\theta)$:
\begin{align}\label{Lemma3}
\sup_{\theta \in \nu}|\check{l}_K(\theta)-E[l_{K0}(\theta)]|=o_p(1).
\end{align}
This can be shown as follows:
$$\check{l}_K(\theta)-E[l_{K0}(\theta)]=l_{K0}(\theta)-E[l_{K0}(\theta)]+\zeta_K(\theta;\boldsymbol{\beta}_0)(\check{\boldsymbol{\beta}}-\boldsymbol{\beta}_0)
+\int_0^\tau \zeta_K(\theta;\Lambda_0)(t)d(\check{\Lambda}-\Lambda_0)(t) + R.$$
Due to the law of large numbers, for fixed $\theta$,
\begin{align}\label{LLN1}
l_{K0}(\theta)-E[l_{K0}(\theta)] \overset {p}{\rightarrow} 0.
\end{align}
Since $||\zeta_K(\theta;\boldsymbol{\beta})||$ is bounded, say $||\zeta_K(\theta;\boldsymbol{\beta})|| \leq M_1$, we have
\begin{align}\label{lemma4deel2}
\sup_{\theta \in \nu}\left|\zeta_K(\theta;\boldsymbol{\beta}_0)(\check{\boldsymbol{\beta}}-\boldsymbol{\beta}_0)\right| \leq M_1 ||\check{\boldsymbol{\beta}}-\boldsymbol{\beta}_0||.
\end{align}
Since $||\zeta_K(\theta;\Lambda)(u)||_\infty$ is bounded, say $||\zeta_K(\theta;\Lambda)(u)||_\infty \leq M_2$, we have
\begin{align}\label{lemma4deel3}
\sup_{\theta \in \nu}\left|\int_0^\tau \zeta_K(\theta;\Lambda)(t)d(\check{\Lambda}-\Lambda_0)(t)\right| \leq M_2 ||\check{\Lambda}-\Lambda_0||_\infty.
\end{align}
Therefore
$$\sup_{\theta \in \nu}\left|\check{l}_K(\theta)-E[l_{K0}(\theta)] \right| \leq \sup_{\theta \in \nu}\left|l_{K0}(\theta)-E[l_{K0}(\theta)] \right| +M_1||\check{\boldsymbol{\beta}}-\boldsymbol{\beta}_0||+ M_2 ||\check{\Lambda}-\Lambda_0||_\infty + R.$$
Using (\ref{LLN1}), the consistency of $\check{\boldsymbol{\beta}}$, the uniform consistency of $\check{\Lambda}$ and the fact that $R=o_p(1)$, we get
$$\sup_{\theta \in \nu}\left|\check{l}_K(\theta)-E[l_{K0}(\theta)] \right|=o_p(1).$$
Finally, in order to verify that $\check{\theta}$ is consistent, we will need to show that the expected log-likelihood is maximized at the truth:
\begin{align}\label{Lemma4}
E[l_{K0}(\theta)]-E[l_{K0}(\theta_0)]<0.
\end{align}
Due to independence between clusters and the fact that all lower dimensional copulas can be regarded as margins of the highest dimensional copula, the log-likelihood $l_K(\theta)$ can be written as a sum of i.i.d. random variables
$$K^{-1}\sum_{i=1}^K \log L_i (\theta;\boldsymbol{\beta},\Lambda)$$
with
\begin{align*}
L_i&=(-1)^{d_i}\frac{\partial^{d_i}}{\partial\{\delta_{ij}=1\}}S(y_{i1},\dots,y_{i,n_i})\\
&= \left(\prod_{j=1}^{n_i}\left[\frac{1}{\varphi_{\theta}'\left(\varphi_{\theta}^{-1}\left(e^{-\Lambda(y_{ij})}
\right)\right)}\right]^{\delta_{ij}}\right)
\varphi_{\theta}^{(d_i)}\left(\sum_{j=1}^{n_i}\varphi_{\theta}^{-1}\left(e^{-\Lambda(y_{ij})}\right)\right)
\end{align*}
where $\partial\{\delta_{ij}=1\}$ is the set of uncensored individuals in cluster $i$.\\\\
Take $\theta \neq \theta_0.$ The law of large numbers, Jensen's inequality and condition C6 imply that
\begin{align*}
\lim_{K \to \infty} l_{K0}(\theta) - l_{K0}(\theta_0) &= E[l_{K0}(\theta)] - E[l_{K0}(\theta_0)]\\
 &= E\left[K^{-1}\sum_{i=1}^K\log L_i(\theta;\boldsymbol{\beta}_0,\Lambda_0)\right]-E\left[K^{-1}\sum_{i=1}^K\log L_i(\theta_0;\boldsymbol{\beta}_0,\Lambda_0)\right] \\
  &= E\left[\log L_1(\theta;\boldsymbol{\beta}_0,\Lambda_0)-\log L_1(\theta_0;\boldsymbol{\beta}_0,\Lambda_0)\right] \\
 &= E\left[\log \frac{L_1(\theta;\boldsymbol{\beta}_0,\Lambda_0)}{L_1(\theta_0;\boldsymbol{\beta}_0,\Lambda_0)}\right]\\
 & \leq \log E \left[\frac{L_1(\theta;\boldsymbol{\beta}_0,\Lambda_0)}{L_1(\theta_0;\boldsymbol{\beta}_0,\Lambda_0)} \right]\\
 & = \log 1\\
 &=0.
\end{align*}
The before last equality results from $L_1(\theta;\boldsymbol{\beta}_0,\Lambda_0)$ being the contribution of cluster 1 to the likelihood $L(\theta;\boldsymbol{\beta}_0,\Lambda_0)$, which is the joint density function of $(y_{11},\dots,y_{1,n_1};\delta_{11},\dots,\delta_{1,n_1})$.\\
Since $\check{\theta}$ maximizes $\check{l}_K(\theta)$, (\ref{Lemma3}) implies that
$$0 \leq \check{l}_K(\check{\theta})-\check{l}_K(\theta_0)=\check{l}_K(\check{\theta})-\check{l}_K(\theta_0) + E[l_{K0}(\theta_0)] - E[l_{K0}(\theta_0)] = \check{l}_K(\check{\theta})- E[l_{K0}(\theta_0)] + o_p(1)$$
$$\Downarrow$$
$$E[l_{K0}(\theta_0)] \leq \check{l}_K(\check{\theta}) + o_p(1).$$
Subtract $E[l_{K0}(\check{\theta})]$ from each side of the inequality to write
\begin{align}\label{eq9}
E[l_{K0}(\theta_0)]-E[l_{K0}(\check{\theta})] \leq \check{l}_K(\check{\theta}) - E[l_{K0}(\check{\theta})] + o_p(1)
\leq \sup_{\theta \in \Theta} |\check{l}_K(\theta)-E[l_{K0}(\theta)]|+o_p(1) = o_p(1).
\end{align}
Now take $\theta$ such that $|\theta - \theta_0| \geq \varepsilon$ for any fixed $\varepsilon >0$. By (\ref{Lemma4}) there must exist some $\gamma_{\varepsilon}>0$ such that
$$E[l_{K0}(\check{\theta})] + \gamma_{\varepsilon} < E[l_{K0}(\theta_0)].$$
It follows that
$$P(|\check{\theta}-\theta_0| \geq \varepsilon) \leq P(E[l_{K0}(\check{\theta})]+\gamma_{\varepsilon}<E[l_{K0}(\theta_0)]).$$
Equation (\ref{eq9}) implies that
$$P(E[l_{K0}(\check{\theta})]+\gamma_{\varepsilon}<E[l_{K0}(\theta_0)]) \to 0 \text{ as } K \to \infty.$$
Therefore
$$P(|\check{\theta}-\theta_0| \geq \varepsilon) \to 0 \text{ as } K \to \infty$$
which proves the consistency of $\check{\theta}$.\\\\
\textbf{Proof of Theorem 3.}
Take a first order Taylor series expansion of $\hat{U}_K(\hat{\theta})$ around and $\theta_0$:
\begin{align}
\hat{U}_K(\hat{\theta})=\hat{U}_K(\theta_0)+(\hat{\theta}-\theta_0)\left.\frac{\partial \hat{U}_K}{\partial \theta}\right|_{\theta=\theta^*}
\end{align}
where $\theta^*$ is between $\hat{\theta}$ and $\theta_0$. It must be the case that $\hat{U}_K(\hat{\theta})=0$ since $\hat{\theta}$ was taken to be the maximum of $L(\theta;\check{\boldsymbol{\beta}},\check{\Lambda})$. Therefore
\begin{align}\label{breuk}
\sqrt{K}(\hat{\theta}-\theta_0)=\frac{\sqrt{K}\hat{U}_K(\theta_0)}{-\left.\frac{\partial \hat{U}_K}{\partial \theta}\right|_{\theta=\theta^*}}.
\end{align}
We already showed that $\hat{\theta}$ consistently estimates $\theta_0$, so the law of large numbers implies that
$$\left.\frac{\partial \hat{U}_K}{\partial \theta}\right|_{\theta=\theta^*} \xrightarrow{\tiny{P}} W(\theta_0)=\lim_{K\to \infty}\left.\frac{\partial {U}_K}{\partial \theta}\right|_{\theta=\theta_0}  \quad\text{(Fisher information)}.$$
We will show that the score equation $\hat{U}_K(\theta_0)$ in the numerator of (\ref{breuk}) follows a normal distribution. Hereto we need a Taylor series expansion of $\hat{U}_K(\theta_0)$ around $\boldsymbol{\beta}_0$ and $\Lambda_0$. Because $\Lambda_0$ is an unspecified function, we will use the Hadamard derivative of $U_K(\theta_0)$ w.r.t. $\Lambda$ at $\Gamma-\Lambda \in BV[0,\tau]$.
\begin{align*}
\frac{d}{dt}\left.\left[K^{-1}\frac{\partial\log L(\theta;\boldsymbol{\beta},\Lambda+t(\Gamma - \Lambda))}{\partial \theta}
\right]\right|_{t=0}=\int_0^\tau \xi_K(\theta;\Lambda)(u)d(\Gamma-\Lambda)(u)
\end{align*}
where
\begin{align*}
\xi_K(\theta;\Lambda)(u) &=K^{-1}\sum_{i=1}^{K}\sum_{j=1}^{n_i} D_{ij}^U Y_{ij}(u)\exp[\boldsymbol{\beta}'\boldsymbol{Z}_{ij}(u)]
\end{align*}
and
\begin{align*}
D_{ij}^U = & \left\{\delta_{ij} \frac{\varphi_{\theta}''\left(\varphi_{\theta}^{-1}\left(H_{ij}\right)\right)}
{\varphi_\theta'\left(\varphi_\theta^{-1}\left(H_{ij}\right)\right)}
 \frac{\partial}{\partial \theta}\left[\varphi_{\theta}'\left(\varphi_{\theta}^{-1}\left(H_{ij}\right)\right) \right]^{-1}\right.\\
&\qquad + \delta_{ij} \varphi_{\theta}'\left(\varphi_{\theta}^{-1}\left(H_{ij}\right)\right)  \frac{\partial}{\partial \theta}\left[-\frac{\varphi_{\theta}''\left(\varphi_{\theta}^{-1}\left(H_{ij}\right)\right)}
{\varphi_\theta'\left(\varphi_\theta^{-1}\left(H_{ij}\right)\right)^3}\right]\\
&\qquad - \frac{\varphi_\theta^{(d_i+1)}\left(\sum_{j=1}^{n_i} \varphi_{\theta}^{-1}\left(H_{ij}\right)\right)}
{\left[\varphi_\theta^{(d_i)}\left(\sum_{j=1}^{n_i} \varphi_{\theta}^{-1}\left(H_{ij}\right)\right)\right]^2}
 \frac{1}{\varphi_{\theta}'\left(\varphi_{\theta}^{-1}\left(H_{ij}\right)\right)}
 \frac{\partial}{\partial \theta}\left[\varphi_{\theta}^{(d_i)}\left(\sum_{j=1}^{n_i} \varphi_{\theta}^{-1}\left(H_{ij}\right)\right)\right]\\
&\qquad \left.+\frac{1}{\varphi_{\theta}^{(d_i)}\left(\sum_{j=1}^{n_i} \varphi_{\theta}^{-1}\left(H_{ij}\right)\right)} \frac{\partial}{\partial \theta}\left[\frac{\varphi_\theta^{(d_i+1)}\left(\sum_{j=1}^{n_i} \varphi_{\theta}^{-1}\left(H_{ij}\right)\right)}
{\varphi_{\theta}'\left(\varphi_{\theta}^{-1}\left(H_{ij}\right)\right)}
 \right]\right\}(-H_{ij}).
 \end{align*}
The derivative of $U_K(\theta)$ w.r.t. $\boldsymbol{\beta}$ is given by
\begin{align*}
\xi_K(\theta;\boldsymbol{\beta})&=K^{-1}\sum_{i=1}^{K}\sum_{j=1}^{n_i} D_{ij}^U \int_0^\tau Y_{ij}(u)\boldsymbol{Z}_{ij}(u)\exp[\boldsymbol{\beta}'\boldsymbol{Z}_{ij}(u)]d\Lambda(u).
\end{align*}
We require $||\xi_K(\theta;\Lambda)||_\infty$ and $||\xi_K(\theta;\boldsymbol{\beta})||$ to be bounded. By condition C5, the terms unique to $\xi_K(\theta;\Lambda)$ and $\xi_K(\theta;\boldsymbol{\beta})$, i.e.
$$||Y_{ij}\exp[\boldsymbol{\beta}'\boldsymbol{Z}_{ij}]||_\infty \quad \text{and} \quad \left|\left|\int_0^\tau Y_{ij}(u)\boldsymbol{Z}_{ij}(u)\exp[\boldsymbol{\beta}'\boldsymbol{Z}_{ij(u)}d\Lambda(u)]\right|\right|$$
are bounded. The common term $||D_{ij}^U||_\infty$ is also bounded.\\\\
A Taylor series expansion of $\hat{U}_K(\theta_0)$ around $\boldsymbol{\beta}_0$ and $\Lambda_0$ gives
$$\hat{U}_K(\theta_0)=U_{K0}(\theta_0)+\xi_K(\theta_0;\boldsymbol{\beta}_0)(\check{\boldsymbol{\beta}}-\boldsymbol{\beta}_0)+\int_0^\tau \xi_K(\theta_0;\Lambda_0)(t)d[\check{\Lambda}(t)-\Lambda_0(t)]+G_K,$$
where $G_K$ is the remainder term for the Taylor series. Since $\check{\Lambda}$ is $\sqrt{K}$-consistent it can be shown that $G_K=o_p(K^{-1/2})$.\\\\
Define the pointwise limit of $\xi_K(\theta,\Lambda)(t)$ as $\xi(\theta,\Lambda)(t)$ and denote $\xi(\theta;\boldsymbol{\beta})=E[\xi_K(\theta;\boldsymbol{\beta})]$. Since $||\xi_K(\theta;\Lambda)||_\infty$ and $||\xi_K(\theta;\boldsymbol{\beta})||$ are bounded, $||\xi(\theta;\Lambda)||_\infty$ and $||\xi(\theta;\boldsymbol{\beta})||$ are too. Therefore
\begin{align}\label{eq10}
\sqrt{K}\hat{U}_K(\theta_0)=\sqrt{K}\left(U_{K0}(\theta_0)+\xi(\theta_0;\boldsymbol{\beta}_0)(\check{\boldsymbol{\beta}}-\boldsymbol{\beta}_0)+\int_0^\tau \xi(\theta_0;\Lambda_0)(t)d[\check{\Lambda}(t)-\Lambda_0(t)]\right)+o_p(1).
\end{align}
By \citet{SpiekermanLin1998}
$$ \sqrt{K} (\check{\boldsymbol{\beta}}-\boldsymbol{\beta}_0) \to \boldsymbol{A}^{-1}\sum_{i=1}^K \boldsymbol{w}_{i.}$$
where $\boldsymbol{w}_{i.}$ is the $i^{\text{th}}$ component of the score function for $\boldsymbol{\beta}$ under the independence working assumption, evaluated at $\boldsymbol{\beta}_0$:
\begin{align*}
\boldsymbol{w}_{i.}= \sum_{j=1}^{n_i} \int_{0}^\tau \{\boldsymbol{Z}_{ij}(u)-E(\boldsymbol{\beta}_0,u)\}dM_{ij}(u)
\end{align*}
with 
$$M_{ij}(t)=\delta_{ij}Y_{ij}(t)-\int_0^t Y_{ij}(u)\exp{\left[\boldsymbol{\beta}'_0\boldsymbol{Z}_{ij}(u)\right]}d\Lambda_0(u).$$
They also showed that $$\sqrt{K}(\check{\Lambda}_0(t,\check{\boldsymbol{\beta}})-\Lambda_0(t))\to \mathcal{W}(t)=K^{-1/2}\sum_{i=1}^K \Psi_i(t)$$
where $\mathcal{W}(t)$ is a zero-mean Gaussian process with variance function
$$E\left[\Psi_1(t)^2\right]$$
with
$$\Psi_i(t)=\int_0^t\frac{dM_{i.}(u)}{s^{(0)}(\boldsymbol{\beta}_0,u)}+\boldsymbol{h}^T(t)\boldsymbol{A}^{-1}\boldsymbol{w}_{i.}$$
and $$\boldsymbol{h}(t)=-\int\limits_{0}^t \boldsymbol{e}(\boldsymbol{\beta}_0,u)d\Lambda_0(u).$$
That's why
\begin{align*}
&\sqrt{K}\left(U_{K0}(\theta_0)+\xi(\theta_0;\boldsymbol{\beta}_0)(\check{\boldsymbol{\beta}}-\boldsymbol{\beta}_0)+\int_0^\tau \xi(\theta_0;\Lambda_0)(t)d[\check{\Lambda}(t)-\Lambda_0(t)]\right)\\
&=\sqrt{K}\left(K^{-1}\sum_{i=1}^K \phi_i(\theta_0)+\xi(\theta_0;\boldsymbol{\beta}_0)K^{-1}\boldsymbol{A}^{-1}\sum_{i=1}^K \boldsymbol{w}_{i.}+\int_0^\tau \xi(\theta_0;\Lambda_0)(t)d\left[K^{-1}\sum_{i=1}^K \Psi_i(t)\right]\right)\\
&=K^{-1/2}\sum_{i=1}^K \left(\phi_i(\theta_0)+\xi(\theta_0;\boldsymbol{\beta}_0)\boldsymbol{A}^{-1} \boldsymbol{w}_{i.}+\int_0^\tau \xi(\theta_0;\Lambda_0)(t)d\Psi_i(t)\right)\\
&=K^{-1/2}\sum_{i=1}^K \Xi_i.
\end{align*}
The central limit theorem implies that $\sqrt{K}\hat{U}_K(\theta_0)$ converges to a normally distributed random variable with mean zero and variance equal to the variance of $\Xi_1$.\\\\
Thus we have
\begin{align}
\sqrt{K}(\hat{\theta}-\theta_0)=\frac{\sqrt{K}\hat{U}_K(\theta_0)}{-\left.\frac{\partial \hat{U}_K}{\partial \theta}\right|_{\theta=\theta^*}}
\end{align}
where
$$\sqrt{K}\hat{U}_K(\theta_0) \xrightarrow{\tiny{D}} N(0,\Var(\Xi_1))$$
and
$$\left.\frac{\partial \hat{U}_K}{\partial \theta}\right|_{\theta=\theta^*} \xrightarrow{\tiny{P}} W(\theta_0).$$
By Slutsky's theorem, $\sqrt{K}(\hat{\theta}-\theta_0)$ converges to a normal distribution with mean zero and variance equal to $$\frac{\Var(\Xi_1)}{W(\theta_0)^2}.$$ The variance of $\Xi_1$ (note that $\Var(\Xi_1)=E[\Xi_1^2]$) can be estimated by $K^{-1}\sum_{i=1}^K \hat{\Xi}_i^2$ where $\hat{\Xi}_i$ is obtained from $\Xi_i$ replacing parameter values by their estimators.\\\\
$W(\theta_0)$ can be estimated by the (minus) derivative of the pseudo score function $\hat{U}_K(\theta)$, evaluated in $\hat{\theta}$.
 \end{document}